\documentclass[12pt]{article}
\usepackage{amsmath}
\usepackage{amsfonts}
\usepackage{amssymb}
\usepackage{graphicx}

\begin{document}

\begin{titlepage}
\ \\
\begin{center}
\LARGE
{\bf
N-body-extended  Channel Estimation\\
for Low-Noise Parameters
}
\end{center}
\ \\
\begin{center}
\large{
M. Hotta${}^{\ast}$,
\footnote[1]{hotta@tuhep.phys.tohoku.ac.jp}
T. Karasawa${}^{\S}$,
\footnote[2]{jidai@ims.is.tohoku.ac.jp}
and M. Ozawa${}^{\S}$
\footnote[3]{ozawa@math.is.tohoku.ac.jp}
}\\
\ \\
\ \\
{\it
${}^\ast$
Department of Physics, Faculty of Science, Tohoku University,\\
Sendai, 980-8578, Japan\\
\ \\
${}^{\S}$
Graduate School of Information Sciences, Tohoku University,\\
Sendai, 980-8579, Japan }
\end{center}
\begin{abstract}
The notion of low-noise channels
was recently proposed and analyzed in detail in order to
describe noise-processes driven by
environment [M. Hotta, T. Karasawa and M. Ozawa, Phys.\ Rev.\ {\bf A72}, 052334 (2005)
]. An estimation theory of low-noise
parameters of channels has also been developed.
In this report, we address the low-noise parameter estimation problem for
the $N$-body extension of low-noise channels.
We perturbatively
calculate the Fisher information of the output states in order to
evaluate the lower-bound of the mean-square error of the parameter estimation.
We show that
the maximum of the Fisher information over all  input states
can be attained by a factorized input state in the leading order of the low-noise parameter.
Thus, to achieve optimal estimation, it is not necessary for there to be
entanglement of  the $N$ subsystems, as long as the true low-noise parameter is  sufficiently small.
\end{abstract}
\end{titlepage}

\section{Introduction}

\ \newline

Quantum channels or quantum operations \cite{NC} can be used to describe low-noise
processes of physical systems coupled weakly with the environment, instead of invoking complicated Hamiltonians. We have introduced the notion of
low-noise quantum channels $\Gamma_{\epsilon}$ characterized by one
low-noise parameter $\epsilon$ in an earlier paper \cite%
{hko}. Low-noise quantum channels are very useful for many physical
applications, including relaxation processes driven by a thermal bath,
decoherence of quantum computers, and rare processes in elementary particle
physics.

Determining of the order of the low-noise parameter $\epsilon$ is often
crucial in various fields. For example, in elementary particle
physics, low-noise signals may be generated by some new physical processes
and the source characterization of the low noise can give us information on the structure of the new physics theory \cite{K}.
In developing scalable quantum computers, it is also
important to determine the order of low noise, as it must be eliminated to
maintain quantum coherence.

In Ref.\ \cite{hko}, we present an estimation theory for noise
parameters in low-noise quantum channels, which we briefly review here. A fundamental quantity
in the theory is the Fisher information of the output state of the channel $%
\Gamma _{\epsilon }$. Consider an input state $\rho _{in}$ for the output
state $\rho =\rho _{out}(\epsilon )$, obtained by 
\begin{equation}
\rho _{out}(\epsilon )=\Gamma _{\epsilon }[\rho _{in}].  \label{q1}
\end{equation}%
The symmetric logarithmic derivative $L$ is defined by%
\begin{equation}
\partial _{\epsilon }\rho =\frac{1}{2}\left( L\rho +\rho L\right) ,\
L^{\dagger }=L.
\end{equation}%
The Fisher information of the output state is defined by 
\begin{equation}
J=\text{Tr}[\rho L^{2}].
\end{equation}%
It is well known that the inverse of $J$ is the lower bound of the mean-square
error of unbiased estimators \cite{Hl, Hol}. The maximum of $J$ over all the
input states can be attained from a pure input state \cite{F}. Hence, in later
discussions, we focus on pure input states. The Fisher information of
output states for an ancilla-extension of the low-noise channel $\Gamma
_{\epsilon }\otimes id_{A}$ was calculated, taking account of entanglement
between the original system $S$ and the ancilla system $A$. For a qubit
system, a characteristic parameter was defined associated with a general
low-noise channel, and the channels for which the prior 
entanglement increases the output Fisher information in terms of the range
of that parameter were characterized. The optimal input pure states were discussed for general
low-noise channels $\Gamma _{\epsilon }$. We introduced an enhancement
factor, representing the ratio of the Fisher information of the ancilla-assisted
estimation to that of the original system, and showed that it is always upper
bounded by 3/2.

We define here an $N$-body-extended channel. We take $N$ identical
systems, with the state space given by $\mathcal{H}^{\otimes N}$, where $%
\mathcal{H}$ is the state space of the original system. Suppose that a
quantum channel $\Gamma_{\theta}$ with an unknown parameter $\theta$ acts on
a state of $\mathcal{H}$. The $N$-body extension of $\Gamma_{\theta}$ is
defined by $\Gamma_{\theta}^{\otimes N}$. The estimation problem of $%
\Gamma_{\theta }^{\otimes N}$ is to find the optimal output measurements and
input states. This N-body-extended problem is nontrivial. Using
collective measurements of the composite system and
entangled input states, the problem cannot be simply reduced to the original
estimation problem of $\Gamma_{\theta}$. Solutions for
optimizing the input states can be obtained only by specific models 
\cite{FI} \cite{imai}, and a complete solution to the problem remains to be determined. For unitary
channel estimations \cite{imai}, the optimal input states of the estimation
are pure states strongly entangled among the $N$ subsystems. Compared with
factorized input states, difficult controls are required to set up such
entangled states in real experiments. If optimization by factorized
input states is possible for a specific class of channels, the physical realization of the entangled input states for the
channels is not important. For example, it is known that factorized-input-state
optimization is satisfied for a generalized Pauli channel \cite{FI}.

\bigskip

\bigskip

In this paper, we discuss the estimation theory of an $N$-body extended
low-noise channel $\Gamma _{\epsilon }^{\otimes N}$, using its
ancilla-extension. The ancilla-extended channel is given by $\Gamma
_{\epsilon }^{\otimes N}\otimes id$, where $id$ is the identical channel of
the ancilla system. As pointed out in Ref.\ \cite{hko}, in order to maximize
the output Fisher information, it is sufficient to adopt an ancilla state
space with dimensions equal to that of the object system. Therefore,
we can assume that the ancilla state space is decomposed into $N$ identical
spaces. By taking an ancilla-extended low-noise channel $\Gamma _{\epsilon
}\otimes id_{A}$, as analyzed in Ref.\ \cite{hko}, the channel $\Gamma
_{\epsilon }^{\otimes N}\otimes id$ can be regarded as $(\Gamma _{\epsilon
}\otimes id_{A})^{\otimes N}$. Hence, we concentrate on the analysis of the
estimation of $(\Gamma _{\epsilon }\otimes id_{A})^{\otimes N}$. Figure 1 describes a typical example of the channel with\ $N=2$%
. An entangled input state for the channel is generated from an initial
factorized state by unitary operators $U_1$, $U_2$, and $V$. After the operation, each of
the non-ancilla subsystems goes through a low-noise channel $%
\Gamma _{\epsilon }$. The obtained output state is measured by
a collective measurement $M$ in order to estimate $\epsilon $. The true
low-noise parameter $\epsilon $ is generally small. Thus, only the analysis in
the leading order of $\epsilon $ is important for applications.
It has been found for the leading order of $\epsilon $ that the maximum value of
the output Fisher information can be attained by a factorized input state
for $(\Gamma _{\epsilon }\otimes id_{A})^{\otimes N}$. This result is
interesting because the number of characteristic parameters of the
low-noise channels is very large. 
\bigskip

\begin{figure}
\centering
\includegraphics[width=9cm,clip]{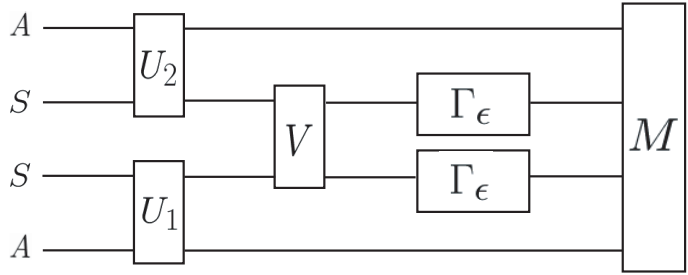}
\caption{{\normalsize Example of }$(\Gamma
_{\epsilon }\otimes id_{A})^{\otimes N}${\normalsize \ with $N=2$. The initial
state is transformed into an entangled state by unitary operators 
$U_1$, $U_2$, and $V$. The non-ancilla
subsystems go through a low-noise channel }$\Gamma
_{\epsilon }${\normalsize . A collective measurement} $M$ 
{\normalsize is made of the output state to estimate the low-noise
parameter }$\epsilon ${\normalsize .}}
\label{hko-pic}
\end{figure}

\bigskip

\bigskip

\section{ Brief Review of Low-Noise Channels}

\ \newline

In this section, we give a brief review of low-noise channels
parametrized by the non-negative low-noise parameter $\epsilon$ \cite{hko}.
The channels are defined in the Kraus representation as

\begin{align}
\Gamma_{\epsilon}[\rho] =\sum_{\alpha}B_{\alpha}(\epsilon) \rho B^{\dagger
}_{\alpha} (\epsilon) +\epsilon\sum_{\beta}C_{\beta}(\epsilon) \rho
C^{\dagger}_{\beta}(\epsilon).  \label{k1}
\end{align}
The Kraus operators must satisfy the following conditions.

\ \newline

(i) The channel is a trace-preserving completely positive (TPCP) map: 
\begin{equation}
\sum_{\alpha}B^{\dagger}_{\alpha}(\epsilon) B_{\alpha}(\epsilon) +\epsilon
\sum_{\beta}C^{\dagger}_{\beta}(\epsilon) C_{\beta}(\epsilon) =\mathbf{1}%
_{S}.  \label{cp}
\end{equation}

\ \newline

(ii)$B_{\alpha}(\epsilon)$ is analytic at $\epsilon=0$, giving the
power series expansion 
\begin{equation}
B_{\alpha}(\epsilon) =\kappa_{\alpha}\mathbf{1}_{S} -\sum^{\infty}_{n=1}
N^{(n)}_{\alpha}\epsilon^{n}  \label{1}
\end{equation}
in the neighborhood of $\epsilon=0$, where $\kappa_{\alpha}$ and $%
N^{(n)}_{\alpha}$ are coefficients and operators, respectively, independent
of $\epsilon$.\newline

\ \newline

(iii) $\kappa_{\alpha}$ satisfies 
\begin{align}
\sum_{\alpha}|\kappa_{\alpha}|^{2} =1.  \label{idlimit}
\end{align}
\ \newline

(iv) $C_{\beta}(\epsilon)$ is analytic at $\epsilon=0$, giving the
power series expansion 
\begin{equation}
C_{\beta}(\epsilon)=M_{\beta}+\sum_{n=1}^{\infty}M_{\beta}^{(n)}\epsilon^{n}
\label{2}
\end{equation}
where $M_{\beta}$ and $M_{\beta}^{(n)}$ are operators independent of $%
\epsilon$.\newline

\ \newline

From condition (iii) (Eq.\ (\ref{idlimit})), the channel automatically
reduces to the identical channel in the noise-vanishing limit: 
\begin{equation}
\lim_{\epsilon\rightarrow+0}\Gamma_{\epsilon}=id_{S}.  \label{icln}
\end{equation}
All physical channels should be TPCP maps and any TPCP map has Kraus
representations. Thus condition (i) naturally applies to low-noise
channels. Conditions (ii) and (iv) simply imply that the channel shows nonsingular
behavior near $\epsilon=0$. 
Therefore, taking proper limits of weak coupling with environment,
 physical processes induced by the environment can be always described by
the low-noise channels.

For general low-noise channels, we can calculate perturbatively the
output Fisher information corresponding to an input state $|\phi\rangle
\langle\phi|$ \cite{hko}. The output Fisher information is evaluated in the
leading order of $\epsilon$ as 
\begin{equation}
J =\frac{1}{\epsilon}\sum_{\beta}\left[ \langle\phi|M^{\dagger}_{\beta
}M_{\beta}|\phi\rangle-\left| \langle\phi|M_{\beta}|\phi\rangle\right| ^{2} %
\right] +O(\epsilon^{0}) ,  \label{f1}
\end{equation}
where $M_{\beta}$ is the lowest-order operator of $C_{\beta}(\epsilon)$
defined by Eq.\ (\ref{2}). Note that this formula can be applied to any
low-noise channel. Hence, in the next section, we may apply this
formula to calculate the output Fisher information of an $N$-body extended
low-noise channel.

We consider an ancillary system $A$ and a composite system $S+A$. The
low-noise channels are trivially extended as $\Gamma_{\epsilon}\otimes
id_{A} $. For an input state $|\Psi\rangle\langle\Psi|$ of $%
\Gamma_{\epsilon}\otimes id_{A}$, the output Fisher information $J_{S+A}$ is
evaluated in the leading order as 
\begin{equation}
J_{S+A}=\frac{1}{\epsilon} \sum_{\beta}\left[ \mathrm{Tr}[\tilde{\rho }%
M^{\dagger}_{\beta}M_{\beta}] -\left| \mathrm{Tr}\left[ \tilde{\rho}
M_{\beta}\right] \right| ^{2} \right] +O(\epsilon^{0}),  \label{sa}
\end{equation}
where $\tilde{\rho}$ is a reduced state of $S$ given by 
\begin{equation}
\tilde{\rho}=\mathrm{Tr}_{A} [|\Psi\rangle\langle\Psi|].
\end{equation}
It should be stressed that $\tilde{\rho}$ can describe any possible
state of the original system $S$.


\section{$(\Gamma_{\protect\epsilon}\otimes id_{A})^{\otimes N}$ Channel
Estimation}

\ \newline

In general, a quantum channel $\Gamma$ possesses a Kraus representation 
\begin{equation}
\Gamma\lbrack\rho]=\sum_{\alpha}A_{\alpha}\rho A_{\alpha}^{\dagger}.
\end{equation}
Using the Kraus representation, $\Gamma^{\otimes N}$ can be written as 
\begin{equation}
\Gamma^{\otimes
N}[\rho^{(N)}]=\sum_{\alpha_{1}\alpha_{2}\cdots\alpha_{N}}A_{\alpha_{1}}%
\otimes A_{\alpha_{2}}\otimes\cdots A_{\alpha_{N}}\rho
^{(N)}A_{\alpha_{1}}^{\dagger}\otimes A_{\alpha_{2}}^{\dagger}\otimes\cdots
A_{\alpha_{N}}^{\dagger}.
\end{equation}
Similarly the Kraus operators of $\Gamma\otimes
id_{A} $ can be derived to be $A_{\alpha}\otimes\mathbf{1}_{A}$. Thus the N-body extension of
the low-noise channel $(\Gamma_{\epsilon}\otimes id_{A})^{\otimes N}$ can be
written as 
\begin{equation}
(\Gamma_{\epsilon}\otimes id_{A})^{\otimes N}[\rho^{(N)}]=\sum_{\alpha
^{\prime}}B_{\alpha^{\prime}}^{(N)}(\epsilon)\rho^{(N)}B_{\alpha^{%
\prime}}^{(N)\dagger}(\epsilon)+\epsilon\sum_{i\beta^{\prime}}C_{i,\beta^{%
\prime}}^{(N)}(\epsilon)\rho^{(N)}C_{i,\beta^{\prime}}^{(N)\dagger}(%
\epsilon),
\end{equation}
where the Kraus operators are expressed by 
\begin{equation}
B_{\alpha^{\prime}}^{(N)}(\epsilon):=B_{\alpha_{1}\cdots\alpha_{N}}^{(N)}(%
\epsilon)=\left( \prod_{i=1}^{N}\kappa_{i}\right) \mathbf{1}^{(N)}\otimes%
\mathbf{1}_{A}^{(N)}+O(\epsilon),
\end{equation}

\begin{equation}
C_{1,\beta^{\prime}}^{(N)}(\epsilon):=C_{1,\beta_{1}\beta_{2}\cdots%
\beta_{N}}^{(N)}(\epsilon)=\left( M_{\beta_{1}}\otimes\mathbf{1}_{A}\right)
\otimes\left( \kappa_{\beta_{2}}\mathbf{1}_{S}\otimes\mathbf{1}_{A}\right)
\otimes\cdots\left( \kappa_{\beta_{N}}\mathbf{1}_{S}\otimes\mathbf{1}%
_{A}\right) +O(\epsilon),  \label{c1}
\end{equation}

\begin{equation}
C_{2,\beta^{\prime}}^{(N)}(\epsilon):=C_{2,\beta_{1}\beta_{2}\cdots%
\beta_{N}}^{(N)}(\epsilon)=\left( \kappa_{\beta_{1}}\mathbf{1}_{S}\otimes 
\mathbf{1}_{A}\right) \otimes\left( M_{\beta_{2}}\otimes\mathbf{1}%
_{A}\right) \otimes\cdots\left( \kappa_{\beta_{N}}\mathbf{1}_{S}\otimes%
\mathbf{1}_{A}\right) +O(\epsilon)  \label{c2}
\end{equation}
and so on.

From Eq.\ (\ref{f1}), the output Fisher information for the
input pure state $|\Psi^{(N)}\rangle\langle\Psi^{(N)}|$ can be evaluated in
the leading order of $\epsilon$ as 
\begin{equation}
J^{(N)} = \frac{1}{\epsilon}\sum_{i=1}^{N} \sum_{\beta^{\prime}} \left[
\langle\Psi^{(N)}|C^{\dagger}_{i,\beta^{\prime}}
C_{i,\beta^{\prime}}|\Psi^{(N)}\rangle- \left| \langle\Psi^{(N)}|
C_{i,\beta^{\prime}}|\Psi ^{(N)}\rangle\right| ^{2} \right] +O(\epsilon^{0}).
\end{equation}
Using $\sum_{\beta_{i}} |\kappa_{\beta_{i}}|^{2} =1$, the expression can
be simplified to 
\begin{equation}
J^{(N)} = \frac{1}{\epsilon}\sum_{i=1}^{N} \sum_{\beta}\left[ \langle
\Psi^{(N)}|M^{\dagger}_{i,\beta} M_{i,\beta}|\Psi^{(N)}\rangle- \left|
\langle\Psi^{(N)}| M_{i,\beta}|\Psi^{(N)}\rangle\right| ^{2} \right]
+O(\epsilon^{0}),
\end{equation}
where the operators $M_{i,\beta}$ are given by 
\begin{equation}
M_{1,\beta} = \left( M_{\alpha}\otimes\mathbf{1}_{A}\right) \otimes (\mathbf{%
1}_{S} \otimes\mathbf{1}_{A})\otimes\cdots(\mathbf{1}_{S} \otimes\mathbf{1}%
_{A}),
\end{equation}
\begin{equation}
M_{2,\beta} = (\mathbf{1}_{S} \otimes\mathbf{1}_{A})\otimes\left( M_{\alpha
}\otimes\mathbf{1}_{A}\right) \otimes\cdots(\mathbf{1}_{S} \otimes \mathbf{1}%
_{A}),
\end{equation}
and so on. Now we define a reduced state $\rho_{i}^{S+A} $ at the $i$%
-th site: 
\begin{equation}
\rho_{i}^{S+A} ={\mathrm{Tr}}_{[i]}[|\Psi^{(N)}\rangle\langle\Psi^{(N)}|],
\end{equation}
where ${\mathrm{Tr}}_{[i]}$ is the trace operation in terms of the $N-1$ sites except
the $i$-th site. The output Fisher information can then be described by
the reduced states as follows: 
\begin{equation}
J^{(N)} = \frac{1}{\epsilon}\sum_{i=1}^{N} \sum_{\beta}\left[ {\mathrm{Tr}}[\rho
_{i}^{S+A} \left( M^{\dagger}_{\beta}\otimes\mathbf{1}_{A}\right) \left(
M_{\beta}\otimes\mathbf{1}_{A}\right) ] - \left| {\mathrm{Tr}}[\rho_{i}^{S+A} \left(
M_{\beta}\otimes\mathbf{1}_{A} \right) ] \right| ^{2} \right]
+O(\epsilon^{0}).
\end{equation}
By performing a trace operation on the ancilla system at the $i$-th
site, we define a reduced state of the $i$-th subsystem as 
\begin{equation}
\tilde{\rho}_{i} ={\mathrm{Tr}}_{A} [\rho_{i}^{S+A}].
\end{equation}
Finally we obtain an expression for the output Fisher information: 
\begin{equation}
J^{(N)} = \frac{1}{\epsilon}\sum_{i=1}^{N} \sum_{\beta}\left[ {\mathrm{Tr}}[\tilde{\rho 
}_{i} M^{\dagger}_{\beta}M_{\beta}] - \left| {\mathrm{Tr}}[\tilde{\rho}_{i} M_{\beta}]
\right| ^{2} \right] +O(\epsilon^{0}).  \label{of}
\end{equation}

Note that Eq.\ (\ref{of}) is expressed by the site-sum ($%
\sum_{i}$) of independent contributions of the output Fisher information in
Eq.\ (\ref{sa}). Hence, by simultaneously maximizing the output
Fisher information at each site, $J^{(N)}$ trivially becomes maximum. We use the optimal input state $%
|\Psi_{opt}\rangle\langle\Psi_{opt}|$ for the single-system channel $%
\Gamma_{\epsilon }\otimes id_{A}$. The state $|\Psi_{opt}\rangle\langle%
\Psi_{opt}|$ maximizes the output Fisher information in Eq.\ (\ref{sa}). 
A factorized input state given by 
\begin{equation}
|\Psi_{opt}^{(N)}\rangle\langle\Psi_{opt}^{(N)}| =(|\Psi_{opt}\rangle
\langle\Psi_{opt}|)^{\otimes N}
\end{equation}
clearly gives maximum $J^{(N)}$. The reduced states $\tilde{\rho }%
_{i,opt}$ are given by 
\begin{equation}
\tilde{\rho}_{i,opt} ={\mathrm{Tr}}_{A} [|\Psi_{opt}\rangle\langle\Psi_{opt}|],
\end{equation}
and maximize each site output Fisher information defined by Eq.\ (\ref{sa}).
The maximum value is given by 
\begin{equation}
J^{(N)}[|\Psi_{opt}^{(N)}\rangle\langle\Psi_{opt}^{(N)}|]
=NJ_{S+A}[|\Psi_{opt}\rangle\langle\Psi_{opt}|],
\end{equation}
where $J_{S+A}[|\Psi_{opt}\rangle\langle\Psi_{opt}|]$ is the output Fisher
information for $\Gamma_{\epsilon}\otimes id_{A}$ with input $%
|\Psi_{opt}\rangle\langle\Psi_{opt}|$. Therefore the prior entanglement of the $%
N $ subsystems cannot increase the Fisher information to larger than $%
NJ_{S+A}[|\Psi_{opt}\rangle\langle\Psi_{opt}|]$.

This result implies that in order to attain an optimal estimation, it is
sufficient to carefully arrange the entanglement between $S$ and $A$, if the state 
$|\Psi_{opt}\rangle$ is an entangled state of $S+A$. It is not necessary to arrange
entanglement of the $N$ subsystems or make collective measurements over the
subsystems.

\ \newline

\ \newline
\ \newline
\ \newline
\textbf{Acknowledgement}\newline

\bigskip

We would like to thank Gen Kimura for useful discussions. This research was
partially supported by the SCOPE project of the MIC and by Grant-in-Aid
for Scientific Research (B)17340021 of the JSPS.

\end{document}